\documentclass[10pt,letterpaper]{article}
\usepackage{opex3}
\usepackage{cite}

\begin{document}

\title{Quantum phase gate for optical qubits with cavity quantum optomechanics }

\author{Muhammad Asjad, Paolo Tombesi, and David Vitali$^*$}

\address{School of Science and Technology, Physics Division, University of Camerino, 62032 Camerino (MC), Italy}

\email{$^*$david.vitali@unicam.it} 



\begin{abstract}
We show that a cavity optomechanical system formed by a mechanical resonator simultaneously coupled to two modes of an optical cavity can be used for the implementation of a deterministic quantum phase gate between optical qubits associated with the two intracavity modes. The scheme is realizable for sufficiently strong single-photon optomechanical coupling in the resolved sideband regime, and is robust against cavity losses.
\end{abstract}



\section{Introduction}
Simple quantum information protocols and quantum gates have been recently implemented with high fidelity in trapped ions and in circuit cavity QED (see e.g., Ref.~\cite{Schoelkopf} for a review). In an efficient quantum network, the information elaborated by a solid state processor at a node should be then robustly encoded in single-photon qubits for long-distance communication and distribution of quantum information. The possibility to implement high-fidelity two-qubit gates between single-photons would greatly facilitate such quantum information routing; an example is provided by perfect Bell-state discrimination for quantum teleportation and entanglement swapping~\cite{Nielsen}, which could be implemented deterministically if a quantum phase gate (QPG) between single photon qubits would be available~\cite{Fortunato}. It is well known that to obtain such a QPG one needs nonlinearities implicit in post-processing measurements or a nonlinear system~\cite{Zeilinger}. A particularly convenient QPG is obtained when the conditional phase shift between the two photonic qubit is equal to $\pi$, because in this latter case the QPG is equivalent, up to local unitary transformations, to a C-NOT gate~\cite{Nielsen,lloyd}. Using all-optical solutions this is not easy to achieve because to process the information one needs strong photon-photon interaction. In fact, to implement quantum information with photons, a nonlinear interaction is needed either to build a two-photon gate operation~\cite{kimble} (but the commonly achievable $\chi^3$ susceptibility factor is typically too small), or the nonlinearity implicit at the detection stage in linear optics quantum computation~\cite{knill}.

Then, one has to think differently as for example using electromagnetically induced transparency (EIT)~\cite{arimondo,Harris1}. Indeed the possibility of reducing the speed of light in atomic media with $\Lambda$-type levels was proposed and experimentally obtained~\cite{Harris2}. Although the two weak fields wave functions traveling with slow group velocity can have a very high nonlinear coupling when they propagate in an atomic media with $\Lambda $-like level configurations~\cite{imamoglu1,imamoglu2},  we need, however, to obtain the same result at the single photon level and this is extremely hard to achieve or even impossible with $\Lambda$-like configurations. A full quantum analysis has shown that a large nonlinear cross-phase shift is achievable using an atomic $M$ level structure \cite{ottaviani1,ottaviani2,rebic}, but it was also shown that a trade-off between the size of the conditional phase shift and the fidelity of the gate exists. This can be avoided in the transient regime, which is however experimentally challenging. More recently, important results have been achieved by exploiting two different solutions able to provide the required effective nonlinearities: i) the strong dispersive coupling of light to strongly interacting atoms in highly excited Rydberg states~\cite{lukin1,fleischhauer1,vuletic1,parigi1,vuletic2,binghe}; ii) a fiber-integrated cavity QED system employing a whispering gallery mode resonator strongly coupled to a single Rubidium atom~\cite{Volz}.

Nevertheless, in recent years it has been proposed~\cite{agarwal}, and then experimentally realized~\cite{kippenberg,teufel,painter0,Karuza}, that EIT-like effects could be obtained also within cavity optomechanical systems. Furthermore, it is well known that the ponderomotive action of light, together with the backaction of the mechanical oscillator interacting with it, is responsible for an effective optical Kerr nonlinearity~\cite{book}, which in turn, may give rise to interesting quantum phenomena, such as squeezing of the cavity output light, as predicted a couple of decades ago \cite{mancini,fabre} and recently experimentally achieved~\cite{brooks,painter,regal}.

In this paper we will show that a QPG for simple photonic qubits with a conditional phase shift equal to $\pi$ is achievable by employing a cavity optomechanical system with sufficiently large single-photon optomechanical coupling, and relatively high mechanical Q-factor $Q_m$. Refs.~\cite{marquardt,stannigel} first suggested that multi-mode optomechanical systems in the single-photon strong coupling regime could be exploited for quantum information processing with photons and phonons, and a first example of optomechanical implementation of a QPG has been recently provided in Ref.~\cite{qpgjpb}. Here we further develop these ideas, by proposing a much simpler scheme, which requires the control of only \emph{two} cavity modes and of a \emph{single} mechanical resonator, rather than four optical cavity modes and two mechanical resonators as in Ref.~\cite{qpgjpb}. Recent progress in the realization of strongly coupled nano-optomechanical systems~\cite{favero,srinivasan,meenehan} suggests that the QPG scheme proposed here could be implemented in the near future.

\section{\bf{The Model} }

We consider an optomechanical system consisting of a mechanical resonator interacting with two optical modes, which is described by the following Hamiltonian
\begin{eqnarray}
\hat{H}=\hbar\omega_1 \hat{a}^\dagger_1 \hat{a}_1+\hbar\omega_2 \hat{a}^\dagger_2 \hat{a}_2+ \hbar\omega_m  \hat{b}^\dagger \hat{b} + \hbar\left(g_1 \hat{a}^\dagger_1 \hat{a}_1+g_2  \hat{a}^\dagger_2 \hat{a}_2 \right)\left(\hat{b}+\hat b^\dagger\right), \label{haml}
\end{eqnarray}
where, $\hat{a}_i (\hat{b})$ and $\hat{a}^\dagger_i (\hat{b}^\dagger)$ are the annihilation and creation operators for the optical cavity (mechanical) modes, with frequency $\omega_{i}/2\pi$ and $\omega_m/2\pi$ respectively, and with $[\hat{a}_i,\hat{a}^\dagger_i]=[\hat{b},\hat{b}^\dagger]=1$; $g_i= (d\omega_{i}/dx)x_{\rm zpf} $ is the $i$-th single-photon optomechanical coupling rate, with $x_{\rm zpf}=\sqrt{\hbar /2m\omega_m}$ the spatial size of the zero-point fluctuation of the mechanical oscillator.

We focus on the simplest choice for an optical qubit, the state space spanned by the lowest Fock states of an optical mode, $|0\rangle $ and $|1\rangle $; to be more specific we want to implement a QPG between the optical qubits associated with two optical cavity modes of the optomechanical system under study. The generic initial (pure) state of the two optical qubits is given by
 \begin{equation}
 |\psi \rangle_{in}= \alpha_{00} |0\rangle_1|0\rangle_2 + \alpha_{01} |0\rangle_1|1\rangle_2 +\alpha_{10} |1\rangle_1|0\rangle_2 + \alpha_{11} |1\rangle_1|1\rangle_2  ,\label{in}
 \end{equation}
corresponding in general to an entangled state of the two modes with up to two photons. A simple proof-of-principle demonstration could be achieved by restricting to \emph{factorized} input states of the two cavity modes, which could be provided by two weak laser pulses driving the two selected cavity modes at frequencies $\omega_{L1}$ and $\omega_{L2}$, similar to the preliminary experimental demonstration of a QPG given in Ref.~\cite{kimble}. In this case the two cavity modes are prepared in a product of two coherent states with amplitudes $\alpha_{d1}$ and $\alpha_{d2}$, $|\psi(0)\rangle = |\alpha_{d1}\rangle_1 |\alpha_{d2}\rangle_2 \simeq \left[|0\rangle_1 +\alpha_{d1}|1\rangle_1\right]\left[|0\rangle_2 +\alpha_{d2}|1\rangle_2\right]$, where the latter expression is valid for $|\alpha_{d1}|,|\alpha_{d2}| \ll 1$. For input laser powers $P_j$, cavity detunings $\Delta_j=\omega_j-\omega_{Lj}$, and decay rates $\kappa_j$, $j=1,2$, the amplitudes are given by $\alpha_{dj}=\sqrt{2 P_j \kappa_j/\left[\hbar \omega_{Lj}\left(\kappa_j^2+\Delta_j^2\right)\right]}$.

It is convenient to move to a frame rotating at the corresponding driving laser frequency for each cavity mode, providing therefore the phase reference for each optical qubit; this is equivalent to move to the interaction picture with respect to the free optical Hamiltonian $H_0 = \hbar\omega_{L1} \hat{a}^\dagger_1 \hat{a}_1+\hbar\omega_{L2} \hat{a}^\dagger_2 \hat{a}_2$, in which the system Hamiltonian becomes
\begin{eqnarray}
\hat{H}=\hbar\Delta_1 \hat{n}_1+\hbar\Delta_2 \hat{n}_2+ \hbar\omega_m  \hat{b}^\dagger \hat{b} + \hbar \omega_m\hat{f}_{\hat{n}}\left(\hat{b}+\hat b^\dagger\right), \label{ham2}
\end{eqnarray}
where we have used the cavity mode photon number operators $\hat{n}_j=\hat{a}^\dagger_j \hat{a}_j$, and we have defined $\hat{f}_{\hat{n}}= \left[g_1 \hat{n}_1+g_2  \hat{n}_2\right]/\omega_m$.

\section{Hamiltonian dynamics}

In order to have a physical description of how the effective optical nonlinearity provided by the optomechanical interaction allows to implement the QPG, we first study the ideal case with no optical and mechanical losses, in which the dynamics is determined solely by the Hamiltonian of Eq.~(\ref{ham2}), i.e., by the unitary operator $ \hat {U}(t)= e^{-i\hat{H} t/\hbar}$. In such a case the dynamics can be exactly solved: in fact, profiting from the fact that both photon number operators $\hat{n}_j$ are conserved, and moving to a photon-number-dependent displaced frame for the mechanical resonator, one can rewrite the unitary evolution operator in a form in which the optical and mechanical evolution operators are conveniently factorized. In fact, acting with the photon number conserving mechanical displacement operator
\begin{equation}\label{displ}
   \hat{D}\left(\hat{f}_{\hat{n}}\right)=\exp\left[\left(\hat{b}^\dagger-\hat{b}\right)\hat{f}_{\hat{n}}\right],
\end{equation}
which separates the Hamiltonian according to
\begin{equation}\label{dispprop}
    \hat{D}\left(\hat{f}_{\hat{n}}\right)\hat{H}\hat{D}^\dagger\left(\hat{f}_{\hat{n}}\right)=\hat{H}_{\rm opt}+\hat{H}_b,
\end{equation}
where
\begin{eqnarray}\label{list1}
   \hat{H}_{\rm opt}&=&\hbar\Delta_1\hat{n}_1+\hbar\Delta_2 \hat{n}_2-\hbar \omega_m \hat{f}_{\hat{n}}^2=\hbar\Delta_1\hat{n}_1+\hbar\Delta_2 \hat{n}_2-\hbar \frac{\left(g_1 \hat{n}_1+g_2\hat{n}_2\right)^2}{\omega_m}, \\
   \hat{H}_b&=&\hbar\omega_m  \hat{b}^\dagger \hat{b}, \label{list2}\\
\end{eqnarray}
one gets
\begin{equation}
\hat {U}(t)= \hat {U}_{\rm opt}(t)\hat{D}^\dagger\left(\hat{f}_{\hat{n}}\right)\hat {U}_b(t)\hat{D}\left(\hat{f}_{\hat{n}}\right),
\end{equation}
where $ \hat {U}_{\rm opt}(t)= e^{-i\hat{H}_{\rm opt} t/\hbar}$, and $ \hat {U}_b(t)= e^{-i\hat{H}_b t/\hbar}$. Therefore the dynamics of the two optical modes is mostly determined by the effective unitary operator $\hat {U}_{\rm opt}(t)$, possessing either self-Kerr terms $\propto \hat{n}_j^2$ and the cross-Kerr term $-2 \hbar g_1 g_2 t\hat{n}_1\hat{n}_2/\omega_m$; the other factor $\hat{D}^\dagger\left(\hat{f}_{\hat{n}}\right)\hat {U}_b(t)\hat{D}\left(\hat{f}_{\hat{n}}\right)$ however also affects the optical mode dynamics since it entangles them with the mechanical resonator, by correlating the resonator position with the photon numbers.

The photon number conserving dynamics allows to stay within the logical space described above, i.e., the one spanned by optical Fock states with no more than one photon (and this will remain true even when we will include optical losses). In this case one can always fix the two detunings in order to eliminate completely the effect of the self-Kerr terms. In fact, within this subspace, $\hat{n}_j^2=\hat{n}_j$, and therefore, taking $\Delta_j=g_j^2/\omega_m$, $j=1,2$, one has the effective unitary operator $ \hat {U}_{\rm opt}(t) = \exp\left[2 i g_1 g_2 t\hat{n}_1\hat{n}_2/\omega_m\right]$, yielding a nonlinear conditional phase shift $\phi_{\rm nl}(t)= 2 g_1 g_2 t/\omega_m$ only when each cavity mode has one photon, i.e.,
   \begin{equation}\label{ideal}
  |0\rangle_1|0\rangle_2 \rightarrow |0\rangle_1|0\rangle_2 ;\,\,\, |0\rangle_1|1\rangle_2 \rightarrow |0\rangle_1|1\rangle_2 ;\,\,\, |1\rangle_1|0\rangle_2 \rightarrow |1\rangle_1|0\rangle_2;\,\,\, |1\rangle_1|1\rangle_2 \rightarrow e^{i\frac{2 g_1 g_2 t}{\omega_m}}|1\rangle_1|1\rangle_2 .
   \end{equation}
Therefore we expect to get a conditional phase shift equal to $\pi$ when the interaction time $t$ is equal to
\begin{equation}
  t_{\pi} = \frac{\pi \omega_m}{2 g_1 g_2}.\label{t-not}
  \end{equation}
We now evaluate the exact Hamiltonian evolution in order to see to what extent the interaction with the mechanical resonator affects and modifies the ideal QPG dynamics defined by Eqs.~(\ref{ideal}) and (\ref{t-not}). The natural choice for the initial state is the factorized state $\hat{\rho}(0)=|\psi \rangle_{in} \langle\psi | \otimes \hat{\rho}_b^{\rm th}$, where $|\psi \rangle_{in} $ is the generic initial state of Eq.~(\ref{in}), and $\hat{\rho}_b^{\rm th}$ is the thermal equilibrium state of the mechanical resonator, with $\bar{n}$ mean thermal phonons. By renumbering $|0\rangle_1|0\rangle_2 \to |0\rangle$,  $|1\rangle_1|0\rangle_2 \to |1\rangle$, $|0\rangle_1|1\rangle_2 \to |2\rangle$, $|1\rangle_1|1\rangle_2 \to |3\rangle$, we can write the state of the whole system at time $t$ as
\begin{eqnarray}\label{svilu}
    &&\hat{\rho}(t)= \hat {U}(t)|\psi \rangle_{in}\langle\psi | \otimes \hat{\rho}_b^{\rm th}\hat {U}^\dagger(t)\\
    &&=\sum_{k,l=0}^3\alpha_k\alpha_l^* \hat {U}_{\rm opt}(t)|k\rangle \langle l|\hat {U}_{\rm opt}^\dagger(t)\otimes \hat{D}^\dagger\left(f_k\right)\hat {U}_b(t)\hat{D}\left(f_k\right) \hat{\rho}_b^{\rm th} \hat{D}^\dagger\left(f_l\right)\hat {U}_b^\dagger(t)\hat{D}\left(f_l\right), \nonumber
\end{eqnarray}
where
\begin{equation}\label{displ2}
    \hat{D}\left(f_k\right)=\exp\left[f_k\left(\hat{b}^\dagger-\hat{b}\right)\right]
\end{equation}
is now a displacement operator acting only on the mechanical resonator degree of freedom, with a c-number displacement $f_k$, with $f_0=0$, $f_1=g_1/\omega_m$, $f_2=g_2/\omega_m$, $f_3=(g_1+g_2)/\omega_m$. We are interested in the state of the two optical qubits only, and therefore we have to trace over the mechanical resonator. Using the explicit expression of $\hat {U}_{\rm opt}(t)$ (with the choice of detuning specified above), and performing the trace, the reduced state of the optical modes reads
\begin{equation}\label{svilu2}
    \hat{\rho}_{\rm opt}(t)= \sum_{k,l=0}^3 c_{k,l}(t)\alpha_k\alpha_l^* \exp\left[i\frac{2 g_1 g_2 t}{\omega_m}\left(\delta_{k,3}-\delta_{l,3}\right)\right]|k\rangle \langle l| ,
\end{equation}
where $\delta_{k,l}$ is the Kronecker delta, while $c_{k,l}(t)$ is the factor describing the decoherence caused by the interaction with the mechanical resonator and whose explicit expression is given by (see the appendix for its derivation)
\begin{equation}\label{deco}
    c_{k,l}(t)=\exp\left[-\left(f_k-f_l\right)^2\left(1-\cos\omega_m t\right)\left(2\bar{n}+1\right)+i\left(f_l^2-f_k^2\right)\sin\omega_m t\right].
\end{equation}
We quantify the QPG performance with the fidelity relative to the ideal pure target state corresponding to a $\pi$ conditional phase shift, that is
\begin{equation}
 |\psi_{tgt} \rangle= \alpha_{00} |0\rangle_1|0\rangle_2 + \alpha_{01} |0\rangle_1|1\rangle_2 +\alpha_{10} |1\rangle_1|0\rangle_2 + e^{i\pi}\alpha_{11} |1\rangle_1|1\rangle_2 = \sum_{k=0}^3 \alpha_k  e^{i\pi\delta_{k,3}}|k\rangle .\label{tgt}
 \end{equation}
The corresponding fidelity $F(t)$ can be written as
\begin{equation}\label{fide}
    F(t)\equiv \langle \psi_{tgt} |\hat{\rho}_{\rm opt}(t)|\psi_{tgt}\rangle = \sum_{k,l=0}^3 c_{k,l}(t)|\alpha_k|^2 |\alpha_l|^2 \exp\left[i\left(\frac{2 g_1 g_2 t}{\omega_m}-\pi\right)\left(\delta_{k,3}-\delta_{l,3}\right)\right].
\end{equation}
Actually, the QPG performance can be characterized by the \emph{gate fidelity}, which is the average of the above quantity over all possible input states of the two qubits~\cite{poyatos}. Since the averages are given by $\overline{|\alpha_k|^4}=1/8$, $\forall k$, and $\overline{|\alpha_k|^2|\alpha_l|^2}=1/24$, $\forall k\neq l$~\cite{poyatos}, using the explicit expression for $c_{k,l}(t)$ of Eq.~(\ref{deco}) implying in particular that $c_{k,k}(t)=1$ $\forall k$ and that $c_{k,l}(t)=c_{l,k}(t)^*$, and the explicit values of the c-numbers $f_k$, we get
\begin{eqnarray}\label{fidegate}
    &&{\cal F}(t)=\overline{\langle \psi_{tgt} |\hat{\rho}_{\rm opt}(t)|\psi_{tgt}\rangle} = \frac{1}{2}\\
    &&+\frac{1}{12}\left\{\exp\left[-\frac{g_1^2}{\omega_m^2}\left(1-\cos\omega_m t\right)\left(2\bar{n}+1\right)\right]\left[\cos\left(\frac{g_1^2}{\omega_m^2}\sin\omega_m t\right) \right.\right. \nonumber \\
   && \left. \left. +\cos\left(\frac{g_1^2+2g_1g_2}{\omega_m^2}\sin\omega_m t + \pi-\frac{2g_1g_2 t}{\omega_m}\right)\right]\right.\nonumber \\
    &&+ \left.\exp\left[-\frac{g_2^2}{\omega_m^2}\left(1-\cos\omega_m t\right)\left(2\bar{n}+1\right)\right]\left[\cos\left(\frac{g_2^2}{\omega_m^2}\sin\omega_m t\right)\right. \right. \nonumber \\
   &&\left.\left. +\cos\left(\frac{g_2^2+2g_1g_2}{\omega_m^2}\sin\omega_m t + \pi-\frac{2g_1g_2 t}{\omega_m}\right)\right]\right.\nonumber \\
   && +\left.\exp\left[-\left(\frac{g_1-g_2}{\omega_m}\right)^2\left(1-\cos\omega_m t\right)\left(2\bar{n}+1\right)\right]\cos\left(\frac{g_2^2-g_1^2}{\omega_m^2}\sin\omega_m t\right)\right.\nonumber \\
   && +\left.\exp\left[-\left(\frac{g_1+g_2}{\omega_m}\right)^2\left(1-\cos\omega_m t\right)\left(2\bar{n}+1\right)\right]\cos\left[\left(\frac{g_1+g_2}{\omega_m}\right)^2\sin\omega_m t+ \pi-\frac{2g_1g_2 t}{\omega_m}\right ]\right\}.\nonumber
\end{eqnarray}
From Eqs.~(\ref{fide}) one can see that the gate fidelity ${\cal F}$ achieves the ideal value of unity when two conditions are satisfied: i) the conditional phase shift is equal to $\pi$, (or more generally to an odd multiple of $\pi$), $2g_1g_2t/\omega_m = (2m+1)\pi$ (integer $m$); ii) $c_{k,l}(t)=1$, $\forall k,l$. Eq.~(\ref{fidegate}) shows that the latter conditions are achieved for generic nonzero couplings $g_1$ and $g_2$ only after every mechanical oscillation period, i.e., when $\omega_m t = 2 p \pi$, $p=1,2,\ldots$. The QPG will be minimally affected by losses for the shortest interaction time $t_{\pi}$ of Eq.~(\ref{t-not}), and therefore, the ideal conditions for a unit-fidelity QPG with a conditional phase shift equal to $\pi$ are
\begin{equation}\label{idecond}
      \frac{2 g_1 g_2 t_{\pi}}{\omega_m }=\pi \;\;\;\omega_m t_{\pi}=2\pi\;\;\Rightarrow g_1 g_2 = \frac{\omega_m^2}{4}.
\end{equation}
Therefore, when optical and mechanical losses are negligible, one can realize an ideal QPG with a simple optomechanical setup by fixing the interaction time between the mechanical resonator and the cavity modes according to Eq.~(\ref{t-not}), and provided that the single-photon optomechanical couplings can be tuned to the strong coupling condition of Eq.~(\ref{idecond}). The interaction time can be controlled in tunable optomechanical systems in which the interaction can be turned on and off, as it could be done for example in the optomechanical setup of Ref.~\cite{wilson}, where a vibrating nanobeam is coupled to the evanescent field of a whispering gallery mode of a microdisk. It is also relevant to stress that under these conditions the QPG is practically insensitive to the effect of thermal noise acting on the resonator, because at the exact gate duration $t_{\pi}$ the fidelity becomes completely independent upon the mean thermal phonon number $\bar{n}$. Eq.~(\ref{fidegate}) shows that instead the gate fidelity significantly drops for increasing $\bar{n}$ as soon as $t \neq t_{\pi}$.

\section{Dissipative dynamics}

Let us now consider the realistic situation in order to see to what extent optical losses, mechanical damping, and thermal noise affect this ideal gate behavior. In that case the evolution is no more analytically tractable, and we will consider the numerical solution of the master equation for the density matrix of the optomechanical system under study.

Introducing the cavity modes decay $\kappa_i$ ($i= 1, 2$), the mechanical damping $\gamma_m = \omega_m/Q_m$, and the mean thermal phonon number associated with the reservoir of the mechanical resonator, $\bar{n}$, the master equation in the usual Born-Markov approximation can be written as~\cite{walls}
\begin{eqnarray} \label{meq}
&&\frac{d}{dt}\hat{\rho}(t)=\frac{1}{i\hbar}[\hat H,\hat\rho(t)]+\frac{\kappa_1}{2}(2\hat a_1\hat\rho(t)\hat a^\dagger_1-\hat a^\dagger_1\hat a_1\hat\rho(t) -\hat\rho(t)\hat a^\dagger_1\hat a_1)\nonumber \\
&&+\frac{\kappa_2}{2}(2\hat a_2\hat\rho(t)\hat a^\dagger_2-\hat a^\dagger_2\hat a_2\hat\rho(t) -\hat\rho(t)\hat a^\dagger_2\hat a_2)  \\
&&+ \frac{\gamma_m}{2}(\bar{n}+1)(2\hat b\hat\rho(t) \hat b^\dagger-\hat b^\dagger\hat b\hat\rho(t) -\hat\rho(t) \hat b^\dagger\hat b)+\frac{\gamma_m}{2}\bar{n}(2\hat b^\dagger\hat\rho(t)\hat b-\hat b \hat b^\dagger\hat\rho(t)-\hat\rho(t)\hat b \hat b^\dagger) , \nonumber
  \end{eqnarray}
where $\hat H$ is the Hamiltonian in Eq.~(\ref{ham2}).

In Fig.~1 we compare the behavior of the gate fidelity ${\cal F}(t)$ in the absence of damping and losses of Eq.~(\ref{fidegate}) versus the dimensionless interaction time $\omega_m t$, either at $\bar{n}=0$ (red dot-dashed curve), and at $\bar{n}=10$ (full black curve) with the corresponding curves in the presence of optical and mechanical damping processes. These latter curves (the dotted blue line corresponds to $\bar{n}=0$, while the green dashed line to $\bar{n}=10$) are obtained from the numerical solution of the master equation Eq.~(\ref{meq}) in the case $\kappa_1=\kappa_2=10^{-2}\omega_m, Q_m= 10^6$, while we have fixed the couplings according to the ideal strong coupling condition of Eq.~(\ref{idecond}) $g_1=g_2 =\omega_m/2$.

We see, as expected, that in the absence of optical and mechanical losses, ${\cal F}(t)=1$ exactly at the interaction time $t_{\pi}$ of Eq.~(\ref{t-not}), regardless the value of the temperature of the mechanical reservoir. At different times the gate performance is strongly affected by thermal noise; what is relevant is that in the presence of realistic values of mechanical damping and of optical loss rates, this scenario is still maintained, with a limited decrease of the gate fidelity.

\begin{figure}[ht]\label{fig1}
\centering
\includegraphics[width=11cm]{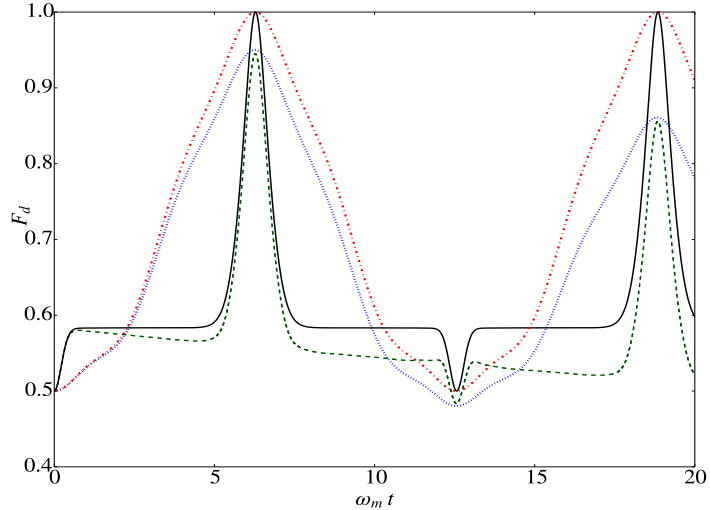}
\caption{Numerical solution of the master equation Eq.~(\protect\ref{meq}) for the gate fidelity ${\cal F}(t)$ versus the dimensionless interaction time $\omega_m t$. We compare four different cases: i) zero damping and losses $\gamma_m=\kappa_1=\kappa_2 = 0$ and $\bar{n}=10$ (full black line); ii) zero damping and losses and $\bar{n}=0$ (red dashed-dotted line); iii) with damping and losses ($\kappa_1=\kappa_2=10^{-2}\omega_m, Q_m= 10^6$,) and $\bar{n}=10$ (green dashed line); iv) with damping and losses ($\kappa_1=\kappa_2=10^{-2}\omega_m, Q_m= 10^6$,) and $\bar{n}=0$ (blue dotted line). The numerical solutions for the zero damping and loss case are indistinguishable from the analytical expression of Eq.~(\protect\ref{fidegate}) either at $\bar{n}=0$ and at $\bar{n}=10$. In all cases we have fixed the couplings according to the ideal strong coupling condition of Eq.~(\protect\ref{idecond}), $g_1=g_2 =\omega_m/2$.}
\end{figure}

\section{Conclusions}

We have proposed a simple optomechanical setup which is able to implement an ideal QPG with a conditional phase shift equal to $\pi$ between two optical qubits associated with the lowest Fock states (zero and one photon) of an optical cavity mode. The scheme is minimal because it employs only two modes of a high-finesse optical cavity and a single mechanical resonator coupled to them. The scheme is robust in the presence of realistic values of optical and mechanical losses and, if the interaction time is appropriately fixed, it is almost completely insensitive to the thermal noise acting on the mechanical resonator. The most stringent and challenging condition is the required strong optomechanical coupling condition, given by Eq.~(\ref{idecond}), in which the single-photon optomechanical coupling must be of the order of the mechanical resonance frequency. Such a condition has not been achieved yet in current solid-state nanomechanical setups, for which record values corresponds to $g_j/\omega_m \sim 10^{-3}$~\cite{favero,srinivasan,meenehan}. On the contrary, such a strong coupling situation is normally achieved in ultracold atom realizations of cavity optomechanics, where the mechanical resonator corresponds to the collective motion of an ensemble of trapped ultracold atoms; for example one has $g/\omega_m \sim 0.3$ in Ref.~\cite{brooks}. The limitation in these latter systems is represented by cavity losses, because in this case one is typically far from the resolved sideband regime $\kappa/\omega_m \ll 1$ which is required here in order that cavity losses do not alter significantly the effective cross-Kerr nonlinear interaction mediated by the resonator, responsible for the QPG dynamics. Therefore the present proposal could be implemented in experimental optomechanical platforms able to combine a significantly large single photon coupling $g/\omega_m \sim 0.5$ with a resolved sideband operation condition $\kappa/\omega_m \ll 1$.

\section{Appendix}

We now derive the explicit expression for the decoherence coefficients $c_{k,l}(t)$ of Eq.~(\ref{deco}). From Eqs.~(\ref{svilu})-(\ref{svilu2}) and using the cyclic property of the trace, one has
\begin{equation}\label{supertrace}
   c_{k,l}(t)={\rm Tr}_b \left[\hat{D}^\dagger\left(f_l\right)\hat {U}_b^\dagger(t)\hat{D}\left(f_l\right)\hat{D}^\dagger\left(f_k\right)\hat {U}_b(t)\hat{D}\left(f_k\right) \hat{\rho}_b^{\rm th}\right],
\end{equation}
which is a thermal average of a combination of displacement operators. We notice then that the factor $\hat {U}_b^\dagger(t)\hat{D}\left(f_l\right)\hat{D}^\dagger\left(f_k\right)\hat {U}_b(t)$ within the trace is just the Heisenberg time evolution for a time $t$ of a displacement operator of a free mechanical resonator, so that
\begin{equation}\label{supertrace2}
   \hat {U}_b^\dagger(t)\hat{D}\left(f_l\right)\hat{D}^\dagger\left(f_k\right)\hat {U}_b(t)=\exp\left[\left(\hat{b}^\dagger e^{i\omega_m t}-\hat{b}e^{-i\omega_m t}\right)\left(f_l-f_k\right)\right].
\end{equation}
Inserting this solution within Eq.~(\ref{supertrace}), and using the property of the displacement operator $\hat{D}(\alpha)\hat{D}(\beta)=\hat{D}(\alpha+\beta)\exp\left[i{\rm Im}(\alpha \beta^*)\right]$, one gets
\begin{equation}\label{supertrace3}
   c_{k,l}(t)={\rm Tr}_b \left\{\hat{D}\left[(f_l-f_k)(e^{i\omega_m t}-1)\right]\hat{\rho}_b^{\rm th}\right\}\exp\left[i(f_l^2-f_k^2)\sin\omega_m t\right].
\end{equation}
Performing the thermal average of the final displacement operator, according to
\begin{equation}\label{thermave}
   \langle \exp\left[\alpha \hat{b}^\dagger -\alpha^*\hat{b}\right]\rangle_{\rm th}=\exp\left[-|\alpha|^2 \left(\bar{n}+\frac{1}{2}\right)\right],
\end{equation}
we finally get Eq.~(\ref{deco}).

\section*{Acknowledgments}
This work has been supported by the European Commission (ITN-Marie Curie project cQOM and FET-Open Project iQUOEMS), and by MIUR (PRIN 2011).

\end{document}